\date{}
\documentclass[11pt]{article}
\usepackage[all]{xy}

\usepackage{amsmath,amsthm,amsfonts,amssymb,amsopn,amscd}     
\usepackage{color}

\DeclareMathOperator{\Tr}{Tr}

\DeclareMathOperator{\Ind}{Ind}

\def\qed{{\unskip\nobreak\hfil\penalty50
\hskip2em\hbox{}\nobreak\hfil$\square$
\parfillskip=0pt \finalhyphendemerits=0\par}\medskip}
\def\proof{\trivlist \item[\hskip \labelsep{\bf Proof.\ }]}
\def\endproof{\null\hfill\qed\endtrivlist\noindent}

\def\a{\alpha}
\def\b{\beta}

\def\e{\varepsilon}
\def\g{\gamma}

\def\l{\lambda}

\def\r{\rho}

\def\th{\theta}
\def\om{\omega}

\newcommand{\ben}{\begin{equation}}
\newcommand{\een}{\end{equation}}

\def\A{{\cal A}}

\def\M{{\cal M}}
\def\N{{\cal N}}

\def\H{{\cal H}}

\def\S{{\cal S}}

\def\f{{\varphi}}
\def\s{{\sigma}}

\def\PSL{{{\rm PSL}(2,\mathbb R)}}

\def\S2{S^{1(2)}}

\def\oin{\om_{\rm in}}
\def\oout{\om_{\rm out}}
\newtheorem{theorem}{Theorem}[section]
\newtheorem{lemma}[theorem]{Lemma}

\newtheorem{proposition}[theorem]{Proposition}

\theoremstyle{definition} 

\theoremstyle{remark} 

\def\PSL{PSU(1,1)}

\def\RR{{\mathbb R}}

\def\SL2{{{\rm SL}(2,\RR)}}

\def\PSL2{{{\rm PSL}(2,\RR)}}

\def\U1{{{\rm V}(1)}}
\def\SU2{{{\rm SV}(2)}}

\def\SU{{{\rm SU}}}

\def\A{{\mathcal A}}

\def\H{{\mathcal H}}

\def\M{{\mathcal M}}
\def\N{{\mathcal N}}

\title{\Huge{Comment on the Bekenstein bound}}
\author{{\sc Roberto Longo}\thanks{Supported in part by the ERC Advanced Grant 669240 QUEST ``Quantum Algebraic Structures and Models'', MIUR FARE R16X5RB55W  QUEST-NET, GNAMPA-INdAM and Alexander von Humboldt Foundation.}
\\ Dipartimento di Matematica, Universit\`a di Roma Tor Vergata \\ 
Via della Ricerca Scientifica, 1 - 00133 Roma, Italy
\\ Email 
{\tt longo@mat.uniroma2.it} \\ [4mm] {\sc Feng Xu} \\
Department of Mathematics, University of California at Riverside\\
Riverside, CA 92521\\
E-mail: {\tt xufeng@math.ucr.edu}
\\[5mm]{\sl Dedicated to Alain Connes on the occasion of his 70th birthday}  
} 

\date{}
\begin{document} 

\maketitle

\begin{abstract}
We propose a rigorous derivation of the Bekenstein upper limit for the entropy/infor\-mation that can be contained by a physical system in a  given finite region of space with given finite energy. The starting point is the observation that the derivation of such a bound provided by Casini \cite{Ca} is similar to the description of the black hole incremental free energy that had been given in \cite{L97}. The approach here is different but close in the spirit to \cite{Ca}. 
Our bound is obtained by operator algebraic methods, in particular Connes' bimodules, Tomita-Takesaki modular theory and Jones' index are essential ingredients inasmuch as the von Neumann algebras in question are typically of type $III$. We rely on the general mathematical framework, recently set up in \cite{L18}, concerning quantum information of infinite systems.  
\end{abstract}

\newpage
\section{Introduction}
\label{sec:intro}
The Bekenstein bound is an universal limit on the entropy that can be contained in a physical system of given size and total energy. If
a system of total energy $E$, including rest mass, is enclosed in a sphere of radius $R$, then the entropy $S$ of the system is bounded by
\[
S\leq \lambda RE\ ,
\]
where $\lambda >0$ is a constant (the value $\lambda = 2\pi$ is often proposed). 

In \cite{Ca}, H. Casini gave an interesting derivation for this bound, based on relative entropy considerations. 
It was observed, following \cite{MMR}, that, in order to get a finite measure for the entropy carried by the system in a region of the space,  one should subtract from the bare entropy of the local state the entropy corresponding to the vacuum fluctuations, which is entirely due to the localisation. 
A similar subtraction can be done to define a localised form of energy. 

The argument in \cite{Ca} is following. One considers a space region $V$ and the von Neumann algebra $\A(O)$ of the observables localised in the causal envelop $O$ of $V$. The restriction $\r_V$ of a global state $\r$ to $\A(O)$ has formally an entropy given by
von Neumann's entropy
\[
S(\r_V) = -\Tr(\r_V\log \r_V)  \ ,
\]
that is known to be infinite. So one subtracts the vacuum state entropy
\[
S_V = S(\r_V) - S(\r^0_V)
\]
with $\r^0_V$ the density matrix of the restriction of the vacuum state $\r^0$ to $\A(O)$. 

Similarly, if $K$ is the  Hamiltonian for $V$, the one considers the difference of the expectations of $K$ in the given state and in the vacuum state
\ben\label{kv}
K_V = \Tr(\r_V K) -  \Tr(\r^0_V K) \ .
\een
The version of the Bekenstein bound in \cite{Ca} is $S_V \leq K_V$, namely
\ben\label{bineq}
S(\r_V) - S(\r^0_V) \leq \Tr(\r_V K) -  \Tr(\r^0_V K)
\een
which is equivalent to
\[
S(\r_V | \r_V^0) \equiv \Tr\big(\r_V (\log \r_V - \log \r^0_V)\big) \geq 0 \, ,
\]
namely to the positivity of the relative entropy. One is then left to estimate the right hand side of \eqref{bineq}. 
Here the (dimensionless) local Hamiltonian $K$ is defined by $\r^0_V = e^{-K}/\Tr(e^K)$, up to a scalar shifting that does not affect the definition of \eqref{kv}. 

The above argument, thought in terms of a cutoff theory, breaks for general Quantum Field Theory as the local von Neumann algebras $\A(O)$ are not of type $I$; under general assumptions, $\A(O)$ is a factor of type $III$ so no trace $\Tr$ and no density matrix $\r$ is definable. Yet, as is well known, modular theory and Araki's relative entropy $S(\f|\psi)$ are definable in general.  We aim at a different argument, close in the spirit to the above discussion, that makes rigorous sense. 

The point is that the above argument is quite similar to the rigorous description of the black hole incremental free energy and entropy given in \cite{L97} in the general Quantum Field Theory framework. Recently, this work led to a universal formula for the incremental free energy, that can be interpreted in several different contexts. This paper is an illustration of this fact. 

We take the point of view that relative entropy is a primary concept and other entropy quantities should be expressed in terms of relative entropies (cf. also \cite{LX17}). This is the case, for example, for the von Neumann entropy. The von Neumann entropy $S(\f)$ of a state $\f$ of a von Neumann algebra $\M$ may be expressed in terms of the relative entropy:
\[
S(\f) = \sup_{(\f_i)}\sum_i S(\f|\f_i)
\]
where the supremum is taken over all finite families of positive linear functionals $\f_i$ of $\M$ with $\sum_i \f_i = \f$ (see \cite{OP}). Clearly $S(\f)$ cannot be finite unless $\M$ is of type $I$. 

However, rather than tracing back the Bekenstein bound to the positivity of the relative entropy, here we are going to rely on the positivity of the incremental free energy, or conditional entropy, which can be obtained in two possible ways: by the monotonicity of the relative entropy in relations to  Connes-St\o rmer's entropy \cite{CS}, or by linking it to Jones' index \cite{Jo}. 
In this respect, our argument is close to the derivation of the bound in \cite{BlCa}, that relies on the monotonicity of the relative entropy. 

\section{Bound for the entropy}
We now are going to compare two states of a physical system, $\oin$ is a suitable reference state, e.g. the
vacuum in QFT, and $\oout$ is a state that can be reached from $\oin$ by some physically realisable process (quantum channel).
We relate the incremental energy and the entropy. 

\subsection{Mathematical and general setting}
With $\N$, $\M$ be von Neumann algebras, an $\N-\M$ bimodule  
is a Hilbert space $\H$ equipped with a normal representation $\ell$ of $\N$ on $\H$ and a normal anti-representation $r$ of $\M$ on $\H$, and $\ell(n)$ commutes with $r(m)$, for all $n\in\N$, $m\in\M$. For simplicity, here we assume that $\N$ and $\M$ are factors.
A vector $\xi\in\H$ is said to be cyclic for $\H$ if it is cyclic for the von Neumann algebra $\ell(\N)\vee r(\M)$ generated by $\ell(\N)$ and $r(\M)$. 

\begin{proposition}\label{Ha}
Let $\a:\N\to\M$ be a completely positive, normal, unital map and $\om$ a faithful normal state of $\M$. Then there exists an $\N-\M$ bimodule $\H_\a$, with a cyclic vector $\xi_\a\in\H$ and left and right actions $\ell_\a$ and $r_\a$, such that
\ben\label{lr}
(\xi_\a, \ell_\a(n)\xi_\a) = \oout(n)\, ,\quad (\xi_\a ,r_\a(m)\xi_\a) = \oin(m) \, ,
\een
with $\oin \equiv\om$, $\oout \equiv \oin\cdot\a$. 
The pair $(\H_\a,\xi_\a)$ with this property is unique up to unitary equivalence. 

Conversely, given an $\N-\M$ bimodule $\H$ with a cyclic vector $\xi\in\H$, with $\om = (\xi,r(\cdot)\xi)$ faithful state of $\M$, there is a unique completely positive, unital, normal map $\a:\N\to\M$ such that $(\H,\xi)=(\H_\a, \xi_\a)$, the cyclic bimodule associated with $\a$ by $\om$. 
\end{proposition}
\proof
For the construction of $(\H_\a , \xi_\a)$, let $\M$ acts on a Hilbert space with cyclic and separating vector $\xi$ such that $\om(m) = (\xi, m\xi)$.  The GNS representation of the algebraic tensor product $\N\odot\M^o$ ($\M^o$ the opposite algebra of $\M$), associated with the state determined by 
\ben\label{st}
n\odot m^o\mapsto (\xi, \a(n)J_\M m^*J_\M\xi)
\een 
 gives $(\H_\a , \xi_\a)$,
see \cite{L18}. 

Conversely, let $\H$ be a $\N-\M$ bimodule with left/right actions $\ell/r$ be given with cyclic vector $\xi$ and $\om = (\xi,r(\cdot)\xi)$ faithful.
Then define $\a:\N\to\M$ by
\ben\label{da}
 r(\a(n))p =  J_\M\, p\,\ell(n^*)\,p\, J_\M 
\een
with $p\in r(\M)'$ the projection onto $\overline{r(\M)\xi}$ and $J_\M$ the modular conjugation of $r(\M)p ,\xi$. 
In order to show that $(\H, \xi) = (\H_\a, \xi_\a)$, by eq. \eqref{st} we have to show that 
\[
(\xi, \ell(n)r(m)\xi) = (\xi_\a, \ell_\a(n)r_\a(m)\xi_\a)\, ,\quad n\in\N,\, m\in\M\, .
\]
Indeed $\M$ acts on the right on $p\H$, $r_p(m) \equiv r(m)|_{p\H}$, with cyclic and separating vector $\xi$ and we may identify $p\H$ with the identity $\M-\M$ bimodule with left action $\ell_p(m) \equiv  J_\M r_p(m^*) J_\M$. 
We have
\begin{multline*}
(\xi_\a, \ell_\a(n)r_\a(m)\xi_\a) = (\xi, \ell_p(\a(n^*))r_p(m)\xi)
= (\xi, J_\M r_p(\a(n))J_\M r_p(m)\xi)\\ = (\xi, J_\M r(\a(n))pJ_\M r(m)\xi)
= (\xi, p\ell(n)p\, r(m)\xi) = (\xi, \ell(n) r(m)\xi)\, .
\end{multline*}
\endproof
Different normal, faithful initial states $\om$ give unitary equivalent bimodules \cite{L18}. 

Let $\a: \N\to\M$ be a normal, unital completely positive map, $\oin$ a faithful normal state of $\M$ as above and $\H_\a, \xi_\a$ as in Prop. \ref{Ha}.  Suppose that $\a$ is faithful, so $\oout = \oin\cdot\a$ is faithful. The converse of Prop. \ref{Ha}, interchanging left and right actions, gives a completely positive, normal, unital map $\a' :\M\to \N$ such that
\[
\oin(\a(n)m) = \oout(n\a'(m))\, ,
\]
called the transpose of $\a$ w.r.t. $\oin$. Similarly as eq. \eqref{da}, $\a'$ is given by
\ben\label{aa}
\ell_\a(\a'(m))q = J_\N\, q\, r_\a(m^*)\,q\, J_\N\, ,
\een
with $q\in \ell_\a(\N)'$ the orthogonal projection onto $\overline{\ell_\a(\N)\xi_\a}$. 

We shall say that $\a$ is {\it left invertible} w.r.t. $\oin$ is $\a'\a = {\rm id}_\N$. We shall say that a state $\oin$ is {\it full} for $\a$ if 
 $\xi_\a$ is cyclic for $r_\a(\M)$ and $\a$ is left invertible w.r.t. $\oin$. 
\begin{lemma}\label{ea}
Suppose that $\xi_\a$ is cyclic for $r_\a(\M)$. Then
 $\oin$ is full for $\a$ 
 iff there exists a conditional expectation $\e : r_\a(\M)'\to \ell_\a(\N)$ preserving the state $(\xi_\a , \cdot \xi_\a)$. 
In this case $\a'(m) = \e(J_\M r_\a(m) J_\M)$. 
\end{lemma}
\proof

By equations \eqref{da} and \eqref{aa}, with $\ell \equiv \ell_\a$ and $r \equiv r_\a$
 we have
\ben
\a(n) =  r^{-1}\big( J_\M\ell(n^*)J_\M \big)\, ,
\een
thus
\[
\ell\big(\a'\cdot\a(n)\big)q =J_\N q \big(r^{-1}r\big( J_\M\ell(n)J_\M \big)\big)qJ_\N
=J_\N q J_\M\ell(n)J_\M qJ_\N\, .
\]
It follows that $\a'(\a(n)) = n$, for all $n\in\N$, iff $J_\N q  J_\M = q$, namely $J_\N = J_\M q = q J_\M$. This is equivalent to say that the map $\e$ is a conditional expectation by Prop. \ref{ce}. 
\endproof
Let $\f,\psi$ be normal, faithful, positive linear functionals of a von Neumann algebra $\M$ and  $\xi_\psi$ be a vector on the underlying Hilbert space such that $\psi = \om_{\xi_\psi} |_\M$, where $\om_{\xi_\psi} = (\xi_\psi , \cdot \, \xi_\psi)$.

We denote Araki's {\it relative entropy} by $S(\f|\psi)$  \cite{Ara}, see \cite{OP}:
\[
S(\f|\psi) = -(\xi_\psi , \log (d\f/d\psi') \,\xi_\psi)\, ;
\] 
where $\psi'$ is the positive linear functional on $\M'$ given by $\psi' = \om_{\xi_\psi} |_{\M'}$ and $d\f/d\psi' $ is Connes' spatial derivative \cite{C80} w.r.t. $\f$ and $\psi'$. 

Essentially all properties of the relative entropy follow at once from Kosaki's variational expression \cite{K86re}. In particular, 
the relative entropy is non-negative and monotone:
\[
S(\f|\psi)\geq 0\, ,\qquad  S(\f\cdot\a |\psi\cdot\a)\leq S(\f|\psi)\, ,
\] 
with $\a: \N\to\M$ a completely positive, unital, normal map. 

Let now $\om$ be a faithful normal state of $\M$ and  $\a: \N\to\M$ is a completely positive, unital, normal map as above. We set
\[
{\rm H}_\om (\a) \equiv \sup_{(\om_i)} \sum_{i}  S(\om|\om_i) - S(\om\cdot\a |\om_i\cdot\a)\, ,
\]
where the supremum is taken over all finite families of positive linear functionals $\om_i$ of $\M$ with $\sum_i \om_i = \om$. 

The {\it conditional entropy} $H(\a)$ of $\a$ is defined by
\[
{\rm H}(\a) = \inf_\om {\rm H}_\om(\a)\, ,
\]
where the infimum is taken over all full states $\om$ for $\a$. If no full state exists, we put ${\rm H}(\a) = \infty$. Clearly ${\rm H}(\a)\geq 0$ because 
${\rm H}_\om(\a)\geq 0$ by the monotonicity of the relative entropy. 

We say that $\a$ is a {\it quantum channel} if its conditional entropy ${\rm H}(\a)$ is finite. 

Let $\a:\N\to\M$ be a quantum channel, that we assume to be faithful for simplicity.  Let $\H_\a$ the bimodule with cyclic vector $\xi$ and left/right action $\ell/r$ associated with $\a$ and the faithful normal state $\om\equiv\oin$ of $\M$ by Prop. \ref{Ha}. As $\a$ is faithful, the output state $\oout = \oin\cdot\a$ is a faithful normal state of $\N$. 

With $\e: r(\M)'\to \ell(\N)$ the minimal conditional expectation (see refs. in \cite{GL}), 
the left {\it modular operator} $\Delta_{\a,\oin}$ of $\a$ with respect to the initial state $\oin$ is the spatial derivative between the states $\oout\cdot\ell^{-1}\cdot\e$ of $r(\M)'$ and $\oin \cdot r^{-1}$ of $r(\M)$
\begin{equation}\label{delta}
\Delta_{\a,\oin} 
= d(\oout\cdot\ell^{-1}\cdot\e) \big/ d(\oin  \cdot r^{-1})\ ,
\end{equation}
thus $
\Delta_{\a,\oin} = d( \om_\xi\cdot\e ) \big/ d(\om_\xi\vert_{r(\M)} )$,  with $\om_\xi = (\xi,\cdot \,\xi)$. 
Then $\Delta\equiv\Delta_{\a,\oin}$ is a positive, non-singular selfadjoint operator on $\H_\a$ and we have
\ben\label{impl}
\Delta^{it}\ell(n)\Delta^{-it} = \ell(\s_t^{\rm out}(n))\, ,\quad 
\Delta^{it}r(m)\Delta^{-it} =  r(\s^{\rm in}_t(m))\, ,
\een
with $\s^{\rm in/out}$ the modular group of $\om_{\rm in/out}$. 

The right modular operator is
\begin{equation}\label{delta'}
\Delta'_{\a,\oin} 
= d(\oout\cdot\ell^{-1}) \big/ d(\oin  \cdot r^{-1}\cdot\e')\ ,
\end{equation}
with $\e': \ell(\N)'\to r(\M)$ the minimal expectation. 

$\log \Delta$ and $\log \Delta'$ are called the left and right {\it modular Hamiltonian} of $\a$ w.r.t. the initial state $\oin$. We have
\[
\log \Delta' = \log \Delta + {\rm H}(\a)\, .
\]
The {\it physical Hamiltonian} $K$ at inverse temperature $\b = \frac1T > 0$, associated with $\a$ and the initial state $\oin$, is a shifting the modular Hamiltonian with natural functoriality properties, and also rescaled in order to get the $\b$-KMS property. By \cite{L18} and Prop. \ref{Hi}, $K$ is given by
\ben\label{fham}
K = -\b^{-1}\log \Delta - \frac{1}{2}\b^{-1}{\rm H}(\a) =  -\b^{-1}\log \Delta' + \frac{1}{2}\b^{-1}{\rm H}(\a)\, .
\een
$K$ may be considered as a local Hamiltonian associated with $\a$ and the state transfer with input state $\oin$. 

The {\it entropy} $S\equiv S_{\a,\oin}$ of $\a$ is here defined as
\[
S = -(\hat\xi , \log \Delta'\hat\xi)  \, ,
\]
where $\hat\xi$ is a vector representative of the state $\oin  \cdot r^{-1}\cdot\e'$ in $\H_\a$. 
$S$ is thus Araki's relative entropy $S \equiv S(\om_{\xi}|_{\ell(\N)} |\om_{\xi} \cdot \e')$ w.r.t. 
the states $\om_\xi |_{\ell(\N)}$ of $\ell(\N)$ and $\om_\xi\cdot\e'$ of $\ell(\N)'$, with $\xi \equiv \xi_\a$.  
Thus $S\geq 0$. 

The quantity 
\[
E = (\hat\xi, K\hat\xi)
\]
is the {\it relative energy} w.r.t. the states $\oin$ and $\oout$. 

The {\it free energy}\footnote{As explained in the footnote in \cite{L18}, the sign of $F$ depends on the left or right modular Hamiltonian choice. Here we consider the right modular Hamiltonian, so $F\geq 0$.}
is now defined by the relative partition function
\[
F =   -\b^{-1}\log(\hat\xi, e^{-\b K}\hat\xi) \, .
\]
\begin{proposition}
$F$ satisfies the thermodynamical relation
\ben\label{trel}
F = E - TS\, .
\een
\end{proposition}
\proof
Similarly as in \cite{L18}, we have $F = \frac12 \b^{-1}{\rm H}(\a)$, indeed by \eqref{fham} we have
\begin{multline*}
\b F = -\log(\hat\xi, e^{-\b K}\hat\xi) = -\log(\hat\xi, \Delta'\hat\xi) + \frac12 {\rm H}(\a)
= -\log||\Delta^{1/2}_{\xi,\hat\xi}\hat\xi ||^2 + \frac12 {\rm H}(\a)\\ =
-\log||J\Delta^{1/2}_{\xi,\hat\xi}\hat\xi ||^2 + \frac12 {\rm H}(\a)=
-\log||\xi ||^2 + \frac12 {\rm H}(\a)
= \frac12 {\rm H}(\a)\, ,
\end{multline*}
with $J$ the modular conjugation w.r.t. $\xi,\hat\xi$. 

Thus eq. \eqref{trel} follows by evaluating the linear relation between $K$ and $\log \Delta'$ in \eqref{fham} on the vector state given by $\hat\xi$. 
\endproof
We then have the following general version of the Bekenstein bound.  
\begin{proposition}
\ben\label{uf}
 S   \leq \b E\, .
\een
\end{proposition}
\proof
As $F = \frac12 \b^{-1}{\rm H}(\a)$, we have $F \geq 0$
because ${\rm H}(\a) \geq 0$. So the inequality follows from the thermodynamical relation \eqref{trel}. 
\endproof

\subsubsection{Fixing the temperature}

Above we have defined the modular Hamiltonian $\log \Delta$ and the physical Hamiltonian $K$ associated with a quantum channel $\a$, given a faithful normal initial state $\om$. $K$ is obtained by the modular Hamltonian by a {\it shifting} and a  {\it scaling}
\[
-\log \Delta\ \xrightarrow{\rm shifting}\  -\log \Delta - \log d(\a)\ \xrightarrow{\rm scaling}\ \b^{-1}\big(-\log \Delta - \log d(\a)\big) \, ,
\]
(left modular Hamiltonian case). 
Here $\log d(\a) = \frac12 {{\rm H}(\a)}$, half of the conditional entropy of $\a$, and is independent of $\om$.  Indeed ${\rm H}(\a)$ is equal to the logarithm of the Jones index of $\H_\a$  (Prop. \ref{Hi}) and the {\it dimension } $d(\a)$ is a tensor categorical notion \cite{LR}.  
 
Both the unitary one-parameter group generated by $-\log \Delta$ and $-\log \Delta - \log d(\a)$ implement the modular flow \eqref{impl}, a property that is not affected by a scalar shifting of the Hamiltonian. The shift is determined by functoriality properties. So both the modular Hamiltonian and the physical Hamiltonian are intrinsic objects. 

Now, the scaling is not intrinsic; it corresponds to a choice of the parametrisation of the modular group to be matched with some physical dynamics (or evolution). 
The one-parameter group generated by $K =  \b^{-1}\big(-\log \Delta - \log d(\a)\big)$: it is the unique rescaling of 
$-\log \Delta - \log d(\a)$ that satisfies the KMS condition w.r.t. the modular group at inverse temperature $\b$. 
In other words, the rescaling is determined once is given by temperature of the system, which is the case of an equilibrium state for a thermodynamical system. 

In Quantum Field Theory, we may have a spacetime region $O$ such that the modular group $\s_t^\om$ of the local von Neumann algebra $\A(O)$ associated with $O$ has a geometric meaning. Namely there is a geometric flow $\th_s : O\to O$ and a re-parametrisation of $\s_t^\om$ that acts covariantly w.r.t $\th$. Here $\om$ may be the vacuum state or any other state. 
In other words, the world lines of $\th$ in $O$ have a modular origin, yet the modular flow $t\mapsto \s_t^\om$ is to be re-parametrised in order to correspond with the geometric flow $\th$. 

The main and well known illustration of the above situation concerns a Rindler wedge region $O$ of the Minkowski spacetime. The vacuum modular group $\Delta^{-it}$ of $\A(O)$ w.r.t. the vacuum state is here  equal to $U(\b t)$, with $U$ the boost unitary one-parameter group  preserving $O$ with acceleration $a$ and $\b$ the Unruh inverse temperature \cite{BW}. In this case the re-parametrisation of the geometric flow is the rescaling by the inverse temperature $\b = 2\pi/a$.  

In general, the re-parametrisation is not just a scaling. As discussed in \cite{CR}, it is however natural to define locally the inverse temperature by
\[
\b_s = \left\Vert \frac{d\th_s}{ds}\right\Vert\, ,
\]
the Minkowskian length of the tangent vector to the modular orbit. Namely $d\tau = \b_s ds$ with $\tau$ proper time along modular trajectories. 

In the following, we shall use this point of view to fix $\b$. This seems to be a good way to compare locally the inverse temperatures of different modular flows. It also gives here indications about the KMS temperatures, yet the right choice is a matter of discussion and is to be determined on physical grounds. 

\subsection{The bound in QFT}

We now read the above structure in a couple of situations in Quantum Field Theory. Our QFT is given by a net of von Neumann algebras $\A(O)$, on a Hilbert space $\H$, associated with spacetime regions $O$ (see \cite{H}). By locality, the von Neumann algebra $\A(O')$ associated with the causal complement $O'$ of $O$ commute with $\A(O)$. For the moment, our spacetime is general, but we shall consider settings where there is a vector $\Omega\in\H$ with the Reeh-Schlieder property, namely $\Omega$ is cyclic and separating for $\A(O)$ if $O$ and $O'$ have non-empty interiors, and regions with geometric modular actions, namely the modular group associated with $\A(O),\Omega$ acts covariantly w.r.t. a geometric flow of $O$. For example, for a Wightman QFT, this is the case for a wedge region on the Minkowski space time, with $\Omega$ the vacuum vector \cite{BW}.  

In \cite{L97,L01}, the relative entropy between the vacuum state $\om$ and a localised charged state (state obtained by $\om$ and a DHR charge \cite{DHR}) is expressed by the thermodynamical relation \eqref{trel} with the relative energy and the incremental free energy. DHR charges on a curved spatime are discussed in \cite{GLRV}. 
Here, we can deal with the output state obtained by any quantum channel, following the general formula in \cite{L18}. 

\subsubsection{Schwarzschild black hole}

Let $O$ be the Schwarzschild black hole region in the Schwarzschild-Kruskal spacetime of mass $M > 0$, namely the region inside the event horizon (see \cite{W}), and $\N\equiv\A(O)$ the local von Neumann algebra associated with $O$ on the underlying Hilbert space $\H$. We consider the Hartle-Hawking vacuum state $\om$ (the global vacuum)
\[
\om(X) = (\Omega, X\Omega) 
\]
and vacuum vector $\Omega$. 
$\H$ is a $\N-\N$ bimodule, indeed the identity $\N-\N$ bimodule $L^2(\N)$ associated with $\Omega$.  

The modular group of $\A(O)$ associated with $\om$ is geometric and indeed corresponds to the geodesic flow. The KMS Hawking-Unruh temperature is
\[
T = 1/8\pi M = 1/ 4\pi R\, ,
\]
with $R= 2M$ the Schwarzschild radius (cf. \cite{U,Sew}). 
Thus our general formula \eqref{uf} gives here
\begin{proposition}
In the Schwarzschild black hole case as above, we have
\[
S \leq 4\pi R E\, ,
\]
with $S$ the  entropy associated with the Hartle-Hawking state and the output state transferred by a quantum channel, and $E$ the corresponding relative energy. 
\end{proposition}
\subsubsection{Conformal QFT}\label{secconf}
We now consider a Conformal Quantum Field Theory on the Minkowski spacetime of any spacetime dimension. Let $O_R$ be the double cone with basis a radius $R>0$ sphere centered at the origin and $\A(O_R)$ the associated local von Neumann algebra. The modular group of $\A(O_R)$ w.r.t. the vacuum state $\om$ has a geometrical meaning:
\[
\Delta_{O_R}^{-is}  = U\big(\Lambda_{O_R}(2\pi s)\big)\ .
\]
Here $U$ is the covariance unitary representation of the conformal group and $\Lambda_{O_R}$ is a one-parameter group of conformal transformation leaving $O_R$ globally invariant and conjugate to the boost one-parameter group of pure Lorentz transformations \cite{HL}. Clearly we have
\[
\Lambda_{O_R}(s) = \delta_R \cdot \Lambda_{O_1}(s) \cdot \delta_{1/R} \, ,
\]
with $\delta_R$ the dilation by $R$. 
We may compare the proper time at a point $\bf x$ with parameter of the flows 
\[
d\tau = \Big|\Big|\frac{d}{ds} \Lambda_{O_R}(s) {\bf x}\Big|\Big|ds = \Big|\Big|\frac{d}{ds} \delta_R \cdot \Lambda_{O_1}(s) \cdot \delta_{1/R} {\bf x}\Big|\Big|ds
= R\Big|\Big|\frac{d}{ds} \Lambda_{O_1}(s) \frac{\bf x}{R}\Big|\Big|ds
\]
(Minkowskian norm); in particular, in the center $\bf 0$ of the sphere, the proper time $\tau_R$ of the flow $\Lambda_{O_R}$ is $R$ times the proper time of the flow $\Lambda_{O_1}$  (cf.  \cite{MR}). 

Now, the inverse temperature $\b_R = \big|\big|\frac{d}{ds} \Lambda_{O_R}(s) {\bf x}\big|\big|_{s=0}$ in $O_R$ 
is maximal on the time-zero basis of $O_R$, in fact at the origin ${\bf x} = {\bf 0}$. Thus the maximal inverse temperatures $\b_R$ in $O_R$ and $\b_1$ in $O_1$ are related by $\b_R = R\b_1$. 

This leads us to fix the  KMS inverse temperature for $\Lambda_{O_R}$  as $\b_R = R\b_1$. 
One indeed computes that $\b_1 = \pi$, half of the Unruh value, and $\b_R = \pi R$.  

Our general formula \eqref{uf} now gives:
\begin{proposition}
Let $O_R$ be a radius $R>0$ double cone in the Minkowski spacetime of any dimension as above, and $\A(O_R)$ the local von Neumann algebra in a conformal QFT. Then
\[
S \leq \pi RE\, .
\]
with $S$ and $E$ the entropy and  energy associated with any quantum channel by the vacuum state. 
\end{proposition}
\proof
$S\leq \b_R E = \b_1 R E\leq \pi RE$.
\endproof
\subsubsection{Boundary CFT}
The analysis in this section is rather interlocutory, less complete than the previous ones. Yet it shows up new aspects as the temperature depends on the distance from the boundary. 

We consider now a 1+1 dimensional Boundary CFT on the right Minkowski half-plane $x>0$. The net $\A_+$ of von Neumann algebras on the half-plane is associated with a local conformal net $\A$ of von Neumann algebras on the real line (time axis) by
\[
A_+(O) = \A(I_-)\vee\A(I_+)\, ;
\]
Here $I_- , I_+$ are intervals of the time axis at positive distance with $I_+ > I_-$ ($I_+$ is on the future of $I_-$) and 
$O$ is the associated double cone $O = I_- \times I_+ \equiv \{(t,x): t\pm x\in I_\pm\}$. 

More generally, in the rational case, we have to consider a (necessarily finite-index) extension of $\A$. However the following discussion remains  the same. 

There is a natural state with geometric modular action \cite{LMR}, that corresponds to the chiral ``2-interval state" and geometric action of the double covering of the M\"obius group \cite{LX04}. We refer to this state as the ``geometric state". 

With $I_- = (a_1 , b_1)$, $I_+ = (a_2 , b_2)$,
in chiral coordinates $u = x+t$, $v= x-t$, the flow $\th^O_s(u,v) =(u_s,v_s)$ has velocity field $(\partial u_s,\partial v_s)$
given by 
\ben\label{flow}
\partial_s u_s = 2\pi\frac{(u_s -a_1)(u_s - b_1)(u_s -a_2)(u_s - b_2)}
{L u_s^2 - 2Mu_s +N} \, ,
\een
with $L = b_1 - a_1 + b_2 - a_2$, $M = b_1 b_2 - a_1 a_2$, $N = b_2 a_2( b_1 - a_1) + b_1 a_1( b_2 - a_2)$, and similarly for $v_s$ \cite{LMR}. 

Let us fix a double cone $O$ with basis of unit length (say $O$ is Lorentz conjugate to a double cone with basis on the space half-axis with length one).  

With $R >0$, let $O_R$ be the double cone associated with the intervals $R I_- = (R a_1 , R b_1)$, $R I_+ = (R a_2 , R b_2)$, namely $ O_R = \delta_R O$, with $\delta_R$ the dilation by $R$ on the half-plane. Then $\th^{ O_R} = \delta_R\cdot \th^O\cdot \delta_{R^{-1}}$. 

As in Section \ref{secconf}, the maximal inverse temperatures are related by
\[
\beta_{O_R} = R\, \beta_O\, .
\]
By choosing the KMS inverse temperatures equal to the maximal temperature we thus have:
\begin{proposition}
In the above setting, with $S$ and $E$ the entropy and  energy in $O_R$ with respect to the geometric state and a quantum channel, we have
\[
S \leq \l_O R E
\]
where the constant $\l_O$ is equal to $\b_O$. It depends on the distance of $O$ from the boundary via the above  flow \eqref{flow}. 
\end{proposition}

\section{Appendix}
We collect here a couple of mathematical results. 
 
\subsection{Conditional expectations}
Let $\N\subset \M$ be von Neumann algebras on a Hilbert space $\H$ and $\xi\in\H$ a cyclic and separating unit vector for $\M$. Let $q\in \N'$ be the projection onto $\overline{\N\xi}$. 

We may consider the associated modular conjugation $J_\M$ of $(\M,\xi)$ on $\H$ and $J_\N$ of $(\N q, \xi)$ on $q\H$.  We may view $J_\N$ as a anti-linear partial isometry on $\H$ by replacing $J_\N$ with $J_\N q$. The map $\g: \M\to \N$ defined by
\[
\g(m)q = J_\N q J_\M m J_\M q J_\N
\]
is normal, completely positive, unital and preserves the state $\f \equiv (\xi, \cdot \xi)$ on $\M$, cf.  \cite{AC, L84}.  
\begin{proposition}\label{ce}
The following are equivalent:

$(i)$ There exists a conditional expectation $\e$ of $\M$ onto $\N$ preserving $\f$;

$(ii)$ $J_\N = J_\M q = q J_\M$;

$(iii)$ $\g |_\N = {\rm id}$.

\noindent
In this case $\e = \g$. 
\end{proposition}
\proof
Assuming $(i)$,
Takesaki's theorem \cite{T} implies that the modular operator of $(\N,\xi)$ on $q\H$ is the restriction of  the modular operator of $(\M,\xi)$, and this easily entails $(ii)$. The implication $(ii)\Rightarrow (iii)$ is immediate. 

Concerning $(iii)\Rightarrow (i)$, notice that  $\g |_\N = {\rm id}$ implies $\g^2 = \g$, thus $\g$ is an expectation of $\M$ onto $\N$ preserving $\f$. The rest is clear. 
\endproof

\subsection{Entropy and index}\label{entroind}
Let $\N , \M$ be factors. The index $\Ind(\H)$ of an $\N-\M$ bimodule $\H$ is the Jones index $[r(\M)': \ell(\N)]$. The index  $\Ind(\a)$ of a normal, unital, completely positive map $\a:\N\to\M$ is the index of the $\N-\M$ bimodule $\H_\a$, namely $\Ind(\a) \equiv \Ind(\H_\a)$ (see \cite{Jo, K86, L89, L90} and \cite{GL} for the non-factor case).   
The index is related to the conditional entropy.  
\begin{proposition}\label{Hi}
${\rm H}(\a) = \log \Ind(\a)$. 
\end{proposition}
\proof
By Lemma \ref{ea} and notations there, $\e = \a'\cdot j$ with $j$ an anti-isomorphism that preserves the state $\om = (\xi, \cdot\xi)$. This easily implies that ${\rm H}_\om(\a') = {\rm H}_\om(\e)$. On the other hand Connes-St\o rmer's entropy ${\rm H}_\om(\e)$ is equal to the logarithm of the index of $\e$ \cite{PP, Hi}, so minimising ${\rm H}_\om(\a')$ over all full states for $\a$ gives the logarithm of the minimal index $\Ind(\H_\a)$ of $[r_\a(\M)': \ell_\a(\N)]$, namely ${\rm H}(\a') = \log \Ind(\H_\a)$. Since $\Ind(\H_{\a'}) = \Ind(\H_\a)$, we have ${\rm H}(\a') = \log \Ind(\H_{\a'})$, hence the proposition follows by replacing $\a'$ with $\a$. 
\endproof

\bigskip

\noindent
{\bf Acknowledgements.} 
We thanks Y. Kawahigashi for the invitation at the Seasonal Institute of the Mathematical Society of Japan ``Operator Algebras and Mathematical Physics'' Tohoku University, Sendai, August 2016, where our collaboration started. 
We are grateful to H. Casini and K.H. Rehren for helpful comments. 


\begin{thebibliography}{99} \itemsep-.2mm
{\small

\bibitem{AC} {\sc L. Accardi, C. Cecchini},
{\it Conditional expectations in von Neumann algebras and a theorem of Takesaki},
J.  Funct.  Anal.  45, no. 2, (1982), 245--273. 

\bibitem{Ara} {\sc H. Araki},
 {\it Relative Hamiltonians for faithful normal states of a von Neumann algebra}, 
 Pub. R.I.M.S., Kyoto Univ. 9 (1973), 165-209.
 
\bibitem{Be}  {\sc J. D. Bekenstein},
{\it Universal upper bound on the entropy-to-energy ratio for bounded systems}, 
Phys. Rev. D 23 (1981), 287. 

\bibitem{BW}  {\sc J. Bisognano, E. Wichmann}, 
{\it On the duality condition for a Hermitean scalar field}, 
J. Math. Phys. 16 (1975), 985.

\bibitem{BlCa} {\sc D. D. Blanco, H. Casini},
{\it Localization of negative energy and the Bekenstein bound},
Phys. Rev. Lett. 111, (2013), 221601.

\bibitem{Ca} {\sc H. Casini},
{\it Relative entropy and the Bekenstein bound}, 
Class. Quantum Grav. 25 (2008), 205021. 

\bibitem{C80} {\sc A. Connes}, 
{\it On the spatial theory of von Neumann algebras}, 
J. Funct. Anal. 35 (1980), no. 2, 153-164. 

\bibitem{C} {\sc A. Connes}, 
``Noncommutative Geometry'' Academic Press, 1994.

\bibitem{CR} {\sc A. Connes, C. Rovelli},
{\it von Neumann algebra automorphisms and time-thermodynamics relation in generally covariant quantum theories}, 
Class. Quant. Grav. 11 (1994), no. 12, 2899--2917. 

\bibitem{CS} {\sc A. Connes, E. St\o rmer},
{\it Entropy for automorphisms of $II_1$ von Neumann algebras},  
Acta Math. 134, 288--306 (1975). 

\bibitem{DHR} {\sc S. Doplicher, R. Haag, J.E. Roberts}, 
{\it Local observables and particle statistics.~I}, 
Comm. Math. Phys. 23 (1971), 199--230.

\bibitem{GL} {\sc L. Giorgetti, R. Longo}, 
{\it Minimal index and dimension for $2- C^*$-categories with finite-dimensional centers},
in preparation.

\bibitem{GLRV} {\sc D. Guido, R. Longo, J.E. Roberts, R.  Verch},
{\it Charged sectors, spin and statistics in quantum field theory on curved spacetimes}, 
Rev. Math.  Phys. 13 (2001), 125--198   

\bibitem{H} {\sc R. Haag}, 
``Local Quantum Physics -- Fields, Particles, Algebras'', 
2nd edn., Springer, New York, 1996.

\bibitem{Hi} {\sc F.  Hiai},
 {\it Minimum index for subfactors and entropy. II},  
 J. Math. Soc. Japan  43, (1991), 673--678. 

\bibitem{HL} {\sc P.D. Hislop, R. Longo}, 
{\it Modular structure of the local algebras associated with the free massless scalar field theory}, 
Comm. Math. Phys. 84, (1982), 71--85. 

\bibitem{Jo} {\sc V. F. R. Jones},
{\it Index for subfactors}, Invent. Math. 72 (1983), 1--25.

\bibitem{K86} {\sc H. Kosaki},
{\it Extension of Jones' theory on index to arbitrary factors}, 
J. Funct. Anal. 66 (1986), no. 1, 123-140. 

\bibitem{K86re} {\sc H. Kosaki}, 
{\it Relative entropy of states: a variational expression}, 
J. Operator Theory 16 (1986), no. 2, 335--348

\bibitem{L84} {\sc R. Longo},  
{\it Solution of the factorial Stone-Weirstrass conjecture}, Invent.  Math. 76 (1984), 145--155.

\bibitem{L89} {\sc R. Longo}, 
{\it Index of subfactors and statistics of quantum fields. I}, 
Comm. Math. Phys. 126 (1989), no. 2, 217-247. 

\bibitem{L90} {\sc R. Longo}, 
{\it Index of subfactors and statistics of quantum fields. II. Correspondences, braid group statistics and Jones polynomial}, 
Comm. Math. Phys. 130 (1990), no. 2, 285-309. 

\bibitem{L97}{\sc R. Longo}, 
\textit{An analogue of the Kac-Wakimoto formula and black hole conditional entropy}, 
Comm.\ Math.\ Phys.\ 186 (1997), 451--479.

\bibitem{L01} {\sc R. Longo}, 
{\it  Notes for a quantum index theorem}, 
Comm. Math. Phys. 222 (2001), 45-96.

\bibitem{LR} {\sc R. Longo, J.E. Roberts}, 
{\it A theory of dimension}, 
K-Theory 11 (1997), 103-159.

\bibitem{L18} {\sc R. Longo}, 
{\it  On Landauer's principle and bound for infinite systems}, 
Comm. Math. Phys. (in press).

\bibitem{LMR} {\sc R. Longo, P. Martinetti, K.H. Rehren},
{\it Geometric modular action for disjoint intervals and boundary conformal field theory}, 
Rev. Math.  Phys.  22 (2010), 331--354.   

\bibitem{LX04} {\sc R. Longo, F. Xu}, 
{\it  Topological sectors and a dichotomy in conformal field theory}, 
Comm. Math. Phys. 251 (2004), 321--364.  

\bibitem{LX17} {\sc R. Longo, F. Xu}, 
{\it  Relative entropy in CFT}, 
arXiv:1712.07283 [math.OA].

\bibitem{MMR} {\sc D. Marolf, D. Minic, S.F. Ross}, 
{\it Notes on spacetime thermodynamics and the observer-dependence of entropy},
Phys. Rev. D 69 064006 (2004). 

\bibitem{MR} {\sc P. Martinetti, C. Rovelli}, 
{\it Diamond's temperature: Unruh effect for bounded trajectories and thermal time hypothesis}, 
Class. Quant. Grav. 20 (2003), 4919--4932.

\bibitem{OP}{\sc M.  Ohya, D. Petz},
``Quantum entropy and its use", Texts and Monographs in Physics. Springer-Verlag, Berlin, 1993

\bibitem{PP}{\sc M.  Pimsner, S.  Popa}, 
 {\it Entropy and index for subfactors}, 
 Ann. Sci. \'Ecole Norm. Sup. (4) 19 (1986), no. 1, 57--106.

\bibitem{Sew} {\sc  G.L. Sewell},  
{\it Relativity of temperature and the Hawking effect}, 
Phys. Lett. 79 A n. 1, 23 (1980). 

\bibitem{T} {\sc M. Takesaki}, 
 {\it Conditional expectation in von Neumann algebras}, 
 J. Funct. Anal.  9 (1972), 306-321.
 
\bibitem{U} {\sc  W. G. Unruh}, 
{\it Notes on black-hole evaporation}, 
Phys. Rev. D 14 (1976), 870--892.

\bibitem{W} {\sc R. M. Wald},  
``General Relativity'', 
University of Chicago Press, Chicago, IL, 2010. 

}

\end{thebibliography}
\end{document}